\vspace{-20pt}
\documentclass[10pt,final,journal,twocolumn]{IEEEtran}
\vspace{-20pt}
\ifCLASSINFOpdf

\else
\usepackage{algorithm}
\usepackage{booktabs}
\usepackage{setspace}
\usepackage{bm}
\usepackage{array}
\fi
\usepackage{cite}
\usepackage[cmex10]{amsmath}
\usepackage{amsthm}
\usepackage{algorithmic}
\usepackage{stfloats}
\usepackage{multicol,multienum}
\usepackage{multirow}
\usepackage{amssymb}
\usepackage{url}
\usepackage{amsmath}
\usepackage{xcolor}
\usepackage[dvips]{graphicx}
\usepackage{subfigure}
\usepackage{caption}
\DeclareMathSizes{12}{8}{8.4}{6}
\vspace{-20pt}

\newtheorem{lemma}{\textbf{Lemma}}

\begin{document}
\hyphenation{op-tical net-works semi-conduc-tor}

\vspace{-2em}
\title{\LARGE Energy-Efficient Trajectory Design for UAV-Enabled Communication Under Malicious Jamming\vspace{-0.25em}}
\author{Yang Wu, Weiwei~Yang, Xinrong~Guan,
        and Qingqing Wu

        \vspace{-2em}

\thanks{

This work is supported by the Natural Science Foundations of China (No. 61771487 and 61471393). (Corresponding author: Weiwei~Yang).

 Yang Wu, Weiwei~Yang, and Xinrong~Guan are with the College of Communications Engineering, Army Engineering University of PLA, Nanjing, 210007, China (e-mails: wuyang0710@163.com; wwyang1981@163.com; geniusg2017@gmail.com).
 Qingqing Wu is with the State Key Laboratory of Internet of Things for Smart City and Department of Electrical and Computer Engineering, University of Macau, Macao, 999078, China (E-mail: qingqingwu@um.edu.mo).}
}
\vspace{-5pt}
\IEEEpeerreviewmaketitle
\vspace{-5pt}
\maketitle
\vspace{-20pt}
\begin{abstract}
In this letter, we investigate a UAV-enabled communication system, where a UAV is deployed to communicate with the ground node (GN) in the presence of multiple jammers. We aim to maximize the energy efficiency (EE) of the UAV by optimizing its trajectory, subject to the UAV's mobility constraints. However, the formulated problem is difficult to solve due to the non-convex and fractional form of the objective function. Thus, we propose an iterative algorithm based on successive convex approximation (SCA) technique and Dinkelbach's algorithm to solve it. Numerical results show that the proposed algorithm can strike a better balance between the throughput and energy consumption by the optimized trajectory and thus improve the EE significantly as compared to the benchmark algorithms.
\end{abstract}


\IEEEpeerreviewmaketitle

\vspace{-10pt}
\section{Introduction}
\IEEEPARstart{U}{nmanned} aerial vehicle (UAV)-enabled wireless communication has attracted increasing attention recently. Compared to the traditional terrestrial communication, UAV-enabled communication is more likely to have line-of-sight (LoS) channels and provides a new degree of freedom for resource allocation via trajectory optimization, thus bringing significant performance improvement \cite{Zeng2016,WQ,xh}.

However, the limited on-board energy of the UAV is one of the biggest challenges in UAV-enabled communications since the UAV requires much propulsion energy to maintain aloft. In \cite{Zeng2017a}, to maximize the energy efficiency (EE) of a UAV base station (BS) rather than the throughput of it, a fractional programming optimization problem was formulated with an analytical UAV propulsion energy consumption model. The EE was significantly improved with the proposed algorithm by striking an optimal balance between the energy consumption and the throughput. This work was then extended to the UAV-enabled relay communication system in \cite{Xiao}.

On the other hand, the broadcasting nature of the radio propagation makes the UAV communication particularly vulnerable to jamming attacks \cite{wqq}. Once the wireless links are jammed, the communication between the UAV and the remote control ground node (GN) will be degraded or even unavailable. However, conventional anti-jamming techniques mainly focus on the power domain, frequency domain and/or antenna spatial domain, via e.g. increasing transmit power, frequency hopping, and/or receive beamforming. Thanks to the UAV's highly controllable maneuverability, it is also appealing to defend against the wireless jamming attack in the spatial domain. In \cite{Wu2019b}, the throughput between the UAV BSs and the GNs in the presence of jammers was investigated via trajectory optimization. In \cite{Wang2018g}, the UAV-enabled relay communication was further studied. Nevertheless, all these works focused on the spectrum efficiency only and the resulting UAV trajectory generally leads to significant propulsion energy consumption, which however may thus decrease the EE dramatically.

\begin{figure}
\begin{center}
  \includegraphics[width=3in,height=1in,angle=0]{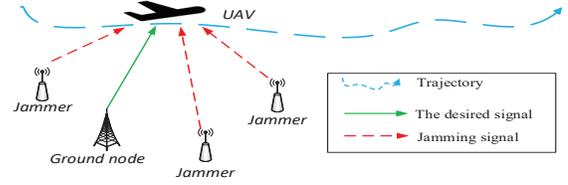}
  \caption{UAV communication under malicious jamming.}\label{system}
\end{center}
\end{figure}

Motivated by the above concerns, in this letter, we study the UAV-enabled communication system in the presence of multiple jammers and formulate an optimization problem that aims to maximize the EE of the UAV. However, the formulated problem is difficult to solve due to the non-convexity caused by the coupled variables and the fraction form of the objective function. To overcome these difficulties, we propose an iterative algorithm to solve it with the aid of successive convex approximation (SCA) technique and Dinkelbach's algorithm. Numerical results show that the proposed algorithm can improve the EE significantly via the proposed trajectory design, especially when the flight time is sufficiently long.
\vspace{-10pt}

\section{System Model and Problem Formulation}

As shown in Fig. 1, we consider a UAV-enabled communication system consisting of a source GN $s$ and a destination UAV $u$ while $M$ jammers are sending jamming signals to the UAV. $s$ and jammer $m \in \left\{ {1,2,...,M} \right\}$ are on the ground with fixed locations. $u$ flies at a fixed altitude $H$, which is the minimum altitude to avoid obstacles. Without loss of generality, 3D Cartesian coordinate system is considered. Thus, the location of $s$, $u$ and $m$ are denoted as ${q_s} = \left\{ {{x_s},{y_s},{0}} \right\}$, ${q_u} = \left\{ {{x_u},{y_u},{H}} \right\}$ and ${q_m} = \left\{ {{x_m},{y_m},{0}} \right\}$, respectively. $u$ is dispatched to fly from a given start point $q_u^{start}$ to an endpoint $q_u^{end}$ to execute the communication task over a finite time period $T$. To facilitate UAV trajectory optimization, $T$ is divided into $N$ equal time slots $d_t$ so that $T\!=\!Nd_t$. Thus, the trajectory of $u$ can be approximately denoted by the combination of discrete locations, i.e., $ {{\bm q}_u[n]= \{ {{x_u}[n],{y_u}[n],H}\}} $. Considering the limited mobility of UAV, with its speed and acceleration speed vector during each time slot $ {{\bm v}_u[n]= \{ {{v_x}[n],{v_y}[n],0}\}} $ and $ {{\bm a}_u[n]= \{ {{a_x}[n],{a_y}[n],0}\}} $, we have the following formulas
\vspace{-6mm}
\begin{spacing}{0.5}
\begin{equation}\label{eq1}
{{\bm q}_u}[n] = {{\bm q}_u}[n - 1] + {{\bm v}_u}[n]d_t + \frac{1}{2}{\bm a}_u[n]{d_t^2},n = 2,3,...,N,
\end{equation}

\begin{equation}\label{eq2}
{{\bm v}_u}[n] = {{\bm v}_u}[n - 1] + {{\bm a}_u}[n]d_t,n = 2,3,...,N,
\end{equation}

\begin{equation}\label{eq3}
{{\bm q}_u}[1] = {\bm q}_u^{start} + {{\bm v}_u}[1]d_t + \frac{1}{2}{{\bm a}_u}[1]{d_t^2},\
\end{equation}

\begin{equation}\label{eq4}
{{\bm q}_u}[N] = {\bm q}_u^{end},\
\end{equation}

\begin{equation}\label{eq5}
\left\| {{{\bm a}_u}[n]} \right\| \le {a_{\max}},\forall n,\
\end{equation}

\begin{equation}\label{eq6}
\left\| {{{\bm v}_u}[n]} \right\| \le {v_{\max}},\forall n,\
\end{equation}

\begin{equation}\label{eq7}
\left\| {{{\bm v}_u}[n]} \right\| \ge {v_{\min}},\forall n,\
\end{equation}
\end{spacing}
\vspace{1mm}
\noindent where ${a_{\max }}$, ${v_{\max }}$ and ${v_{\min }}$ denote the UAV's maximum acceleration speed, maximum flying speed and minimum flying speed, respectively.

The field trial have demonstrated that the air-to-ground (A2G) communication channel between GNs and UAVs are mainly dominated by the large scale path-loss\vspace{-1pt}\footnote{The LoS probability in a rural environment exceeds 95$\%$ for a horizontal ground distance of 2 kilometers when UAV's height is beyond 80 meters.} \cite{x}. Hence, the channel power gains from $s$ to $u$ and $m$ to $u$ can be denoted as ${g_{s,u}}[n] = {\beta _0}{\left\| {{{\bm q}_u}[n] - {{\bm q}_s}} \right\|^{ - 2 }}$ and ${g_{m,u}}[n] = {\beta _0}{\left\| {{{\bm q}_u}[n] - {{\bm q}_m}} \right\|^{ - 2 }}$, respectively, where ${\beta _0}$ is the channel power gain at the reference distance $d_0=1$ m.
Thus, the achievable throughput from $s$ to $u$ at time slot $n$ is given by
\vspace{-1mm}
\begin{equation}\label{eq12}
{R_{s,u}}[n] = B{{\log }_2}\left( {1 + \frac{{{P_s}{g_{s,u}}[n]}}{{ {\sum\limits_{m = 1}^M {P_m}{g_{m,u}}[n] + {\sigma ^2}} }}} \right),\
\end{equation}
\noindent where $B$, ${{P_s}}$, ${{P_m}}$ and ${{\sigma ^2}}$ denote the channel bandwidth, the transmit power of $s$, $m$ and the power of additive white Gaussian noise (AWGN), respectively.

The propulsion energy consumption of a fixed-wing UAV can be expressed as \cite{Zeng2017a}
\vspace{-3mm}
\begin{spacing}{0.7}
\begin{equation}\label{eq15}
{{E_{U\!A\!V\!}}}\! =\! d_t\!\sum\limits_{n = 1}^N \!{\left[ {{c_1}{{\left\| {{{\bm{v}}_u}[n]} \right\|}^3} \!+ \! \frac{{{c_2}}}{{\left\| \!{{{\bm{v}}_u}[n]} \!\right\|}}\left( {\!1 \!+\!\frac{{{{\left\| \!{{{\bm{a}}_u}[n]}\! \right\|}^2}}}{{{g^2}}}}\! \right)} \!\right]}  \!+ \!{\Delta _K},\
\end{equation}
\end{spacing}
\noindent where $c_1$ and $c_2$ are two constant parameters related to aerodynamics, $g$ represents the gravitational acceleration. ${\Delta _K} = \frac{1}{2}J\left( {{{\left\| {{{\bm{v}}_u}[N]} \right\|}^2} - {{\left\| {{{\bm{v}}_u}[1]} \right\|}^2}} \right)\ $ denotes the change of kinetic energy of the UAV, whose value is only related to the UAV's mass $J$ and the initial and final speed.

In this letter, we aim to maximize the EE of the UAV by jointly optimizing its trajectory ${\bm q}_u[n]$, speed ${\bm v}_u[n]$ and acceleration ${\bm a}_u[n]$. The problem can be formulated as

\begin{spacing}{0.5}
\begin{equation}\label{eq16}
\mathop {\max }\limits_{{\bm{q}_u}[n],{\bm{v}_u}[n],{\bm{a}_u}[n]} \frac{{\sum\limits_{n = 1}^N {{R_{s,u}}[n]} }}{{{E_{UAV}}}}\
\end{equation}
\end{spacing}
\begin{equation*}\label{eq17}
\;\;\;\;\;\;\rm{s.t.}\;\;\;\;(\ref{eq1})-(\ref{eq7}).
\end{equation*}

\noindent Problem (\ref{eq16}) is difficult to solve due to the non-convex objective function with a fractional form and the non-convex constraints (\ref{eq7}). In the sequel, an iterative algorithm is proposed to search for a Karush-Kuhn-Tucker (KKT) point of (\ref{eq16}) by leveraging the SCA technique and Dinkelbach's algorithm.

\section{Proposed Algorithm}

\subsection{Reformulation of Objective Function in (\ref{eq16})}

To transform the denominator of the objective function into convex, slack variable $\tau [n]$ is introduced and (\ref{eq16}) is thus reformulated as
\vspace{-2mm}
\begin{subequations}\label{eq17}
\begin{spacing}{0.3}
\begin{flalign}\label{eq17a}
\mathop {\max }\limits_{\scriptstyle{\bm{q}_u}[n],{\bm{v}_u}[n],\hfill\atop
\scriptstyle{\bm{a}_u}[n],\tau [n]\hfill} \frac{{\sum\limits_{n = 1}^N {B{{\log }_2}\left( {1 + \frac{{{P_s}{\beta _0}{{\left\| {{{\bm{q}}_{{u}}}[n] - {{\bm{q}}_s}} \right\|}^{ - 2}}}}{{{\sum\limits_{m = 1}^M  {P_m}{g_{m,u}[n]} + {\sigma ^2}} }}} \right)} }}{{d_t\!\sum\limits_{n = 1}^N\! {\left[ {{c_1}{{\left\| {{{\bm{v}}_u}[n]} \right\|}^3} \!\!+ \!\! \frac{{{c_2}}}{{\tau [n]}}\left( {1\!\! +\! \!\frac{{{{\left\| {{{\bm{a}}_u}[n]} \right\|}^2}}}{{{g^2}}}} \right)} \right]} \!\! + \!\!{\Delta _K}}}\
\end{flalign}
\begin{flalign}\label{eq17b}
\!\!\!\!\!\!\!\!\!\!\!\!\!\!\!\!\!\!\!\!\!\!\!\! {\rm{s.t.}}\;\;\tau [n] \ge {v_{\min }},\forall n,\
\end{flalign}
\begin{flalign}\label{eq17c}
\tau {[n]^2} \le {\left\| {{\bm{v}}_u[n]} \right\|^2},\forall n,\
\end{flalign}
\end{spacing}
\end{subequations}
\vspace{-1mm}
\begin{equation*}\label{eq01}
(\ref{eq1})-(\ref{eq7}).\;\;\;\;\;\;\;\;\;\;\;\;\;\;\;\;\;\;\;\;
\end{equation*}

The equivalence can be verified by contradiction. When problem (\ref{eq16}) and (\ref{eq17}) achieve optimal solution, if (\ref{eq17c}) holds with inequality, we can always improve the value of $\tau[n]$ to increase the value of the objective function. Nevertheless, (\ref{eq17c}) is non-convex.

Meanwhile, to transform the numerator of the objective function into concave, by introducing two slack variables ${{{ L}_{s,u}}[n]}$ and ${{{ I}_u}[n]}$, ${\sum\limits_{n = 1}^N {{R_{s,u}}[n]} }$ is first rewritten as
\vspace{-2mm}
\begin{spacing}{0.9}
\begin{equation}\label{eq22}
\sum\limits_{n = 1}^N {{{\tilde R}_{s,u}}[n]}  = \sum\limits_{n = 1}^N {B{{\log }_2}\left( {1 + \frac{1}{{{L_{s,u}}[n]{I_u}[n]}}} \right)} ,\forall n ,\
\end{equation}
\end{spacing}
\noindent with additional constraints
\begin{spacing}{0.5}
\begin{equation}\label{eq233}
{P_s}{g_{s,u}}[n] \ge {{ L}_{s,u}}{[n]^{ - 1}},\forall n,\:
\end{equation}
\end{spacing}
\noindent and
\vspace{-3mm}
\begin{spacing}{0.7}
\begin{equation}\label{eq244}
\sum\limits_{m = 1}^M{{P_m}{g_{m,u}}[n] + {\sigma ^2}}  \le {{ I}_u}[n],\forall n.\
\end{equation}
\end{spacing}
\noindent The equivalence can be similarly proved by contradiction. Specifically, when the optimal solution is obtained, if (\ref{eq233}) and (\ref{eq244}) hold with inequalities, we can always decrease ${L}_{s,u}{[n]}$ and ${I}_u[n]$ to improve the objective value. However, (\ref{eq244}) is non-convex.

Noting that ${{\tilde R}_{s,u}}[n]$ is convex with respect to ${L}_{s,u}{[n]}$ and ${I}_u[n]$. Then, we use the following lemma to obtain a lower bound of the numerator of the objective function.
\vspace{-1mm}
\begin{spacing}{0.7}
\begin{lemma}\label{lemma2}
For any given feasible point $({ L}_{s,u}^f[j],{{ I}_u^f}[j])$, ${{{\tilde R}_{s,u}}[j]}$ is lower bounded  by
\vspace{-1mm}
\begin{equation}\label{eq26}
\begin{array}{l}
\tilde R_{_{s,u}}^l[j]\  = B{\log _2}(1 + 1/{{L}}_{s,u}^f[j]{{I}}_u^f[j])\\
{\rm{                }} + {A_{s,u}}({{{L}}_{s,u}}[j] - {{L}}_{s,u}^f[j]) + {B_{s,u}}({{{I}}_u}[j] - {{I}}_u^f[j]),
\end{array}
\end{equation}

\noindent where ${A_{s,u}} =  - B{\log _2}e/({{L}}_{s,u}^f[j] + {({{L}}_{s,u}^f[j])^2}{{I}}_u^f[j])\ $ and ${B_{s,u}} =  - B{\log _2}e/({{I}}_u^f[n] + {({{I}}_u^f[n])^2}{{L}}_{s,u}^f[n])$.
\end{lemma}
\end{spacing}
\begin{proof}
Since $f(x,y) = {\log _2}(1 + 1/xy)$ is a convex function, its first-order Taylor expansion provides a global under-estimator at a given feasible point $({x^f},{y^f})$, i.e.,
\vspace{-2mm}
\begin{equation}\label{eq27}
\begin{array}{l}
{\log _2}(1 + 1/xy) \ge {\log _2}(1 + 1/{x^f}{y^f})\\
 - (x - {x^f}){\log _2}e/({x^f} + {({x^f})^2}{y^f})\\
 - (y - {y^f}){\log _2}e/({y^f} + {({y^f})^2}{x^f}).
\end{array}\
\end{equation}

\noindent Thus, by applying $x={{L}_{s,u}}[n]$, $y={I}_u[n]$, \textbf{\emph{Lemma}} \emph{1} is proved.
\end{proof}

\vspace{-10pt}
\subsection{Reformulation of Constraints (\ref{eq7}) and (\ref{eq17c})}

With the first-order Taylor expansion of $V_u^{}[n] = {\left\| {{{\bm v}_u}[n]} \right\|^2}\ $ at feasible point ${\bm v}_f[n]$, i.e., $V_u^l[n] = {\left\| {{\bm{v}}_u^f[n]} \right\|^2} + 2{({\bm{v}}_u^f[n])^T}({{\bm{v}}_u}[n] - {\bm{v}}_u^f[n])$, we square both sides of (\ref{eq7}) as ${\left\| {{{\bm v}_u}[n]} \right\|^2} \ge v_{\min }^2$ and then convert (\ref{eq7}) and (\ref{eq17c}) into convex constraints approximately as
\begin{spacing}{0.5}
\begin{equation}\label{eq30}
{v_{{{\min }}}^2} \le V_u^l[n],\forall n,\
\end{equation}
and
\begin{equation}\label{eq29}
\tau {[n]^2} \le V_u^l[n],\forall n,\
\end{equation}
\end{spacing}
\noindent respectively.

\vspace{-10pt}
\subsection{Reformulation of Constraint (\ref{eq244})}

By introducing slack variable $ d_m[n]$, constraint (\ref{eq244}) can be substituted by
\begin{spacing}{0.5}
\begin{equation}\label{eq311}
\sum\limits_{m = 1}^M {P_{m}\beta_0d_{m}{{[n]}^{ - 1 }} + {\sigma ^2}}  \le I_{u}[n],\forall n,\
\end{equation}

\begin{equation}\label{eq321}
d_m[n] \le {\left\| {\bm{q}_u[n] - \bm{q}_m} \right\|^{ 2}},\forall n,\
\end{equation}

\begin{equation}\label{eq331}
\!\!\!\!\!\!\!\!\!\!\!\!\!\!\!\!\!\!\!\!\!\!\!\!\!\!\!\!\!\!\!\!\!\!d_m[n] \ge 0,\forall n.\
\end{equation}
\end{spacing}

\vspace{1mm}

\noindent The equivalence can also be proved by contradiction. When the optimal solution is obtained, constraint (\ref{eq321}) can hold with equality since otherwise, we can always increase $d_m[n]$ to enhance the value of the objective function.

Then, with the first-order Taylor expansion of ${\left\| {{\bm{q}_u}[n] - {\rm{ }}{\bm{q}_m}} \right\|^2}$, we derive the lower bound of right hand side of constraint (\ref{eq321}) as
\vspace{-1mm}
\begin{spacing}{0.5}
\begin{equation}\label{eq335}
\begin{array}{l}
q_{u,m}^l[n] = 2{x^f}[n]x[n] - {({x^f}[n])^2} + x_m^2 - 2{x_m}x[n]\\
{\rm{         }} + 2{y^f}[n]y[n] - {({y^f}[n])^2} + y_m^2 - 2{y_m}y[n],
\end{array}
\end{equation}

\noindent and transform constraint (\ref{eq321}) as a convex constraint

\begin{equation}\label{eq3210}
{d_m}{\rm{[}}n{\rm{]}} \le q_{u,m}^l[n],\forall n.
\end{equation}
\end{spacing}

\vspace{-10pt}
\subsection{Overall Algorithm and Convergence}

With the derived lower bound of the numerator of objective function $\tilde R_{s,u}^l[n]$ and the derived convex constraints (\ref{eq30}), (\ref{eq29}), (\ref{eq311}), (\ref{eq331}) and (\ref{eq3210}), (\ref{eq16}) can be reformulated as
\begin{spacing}{0.6}
\begin{equation}\label{eq31}
\mathop {\max }\limits_{\bf{\Theta }} \frac{{\sum\limits_{n = 1}^N {\tilde R_{s,u}^l[n]} }}{{{{\tilde E}_{UAV}}}}
\end{equation}
\begin{equation*}\label{eq00}
\;\;\;\;\;\;\rm{s.t.}\;\;\;\;(\ref{eq1})-(\ref{eq6}),(\ref{eq17b}),(\ref{eq233}),(\ref{eq30})-(\ref{eq311}),(\ref{eq331}),(\ref{eq3210}).
\end{equation*}
\end{spacing}
\vspace{1mm}
\noindent where ${\bf{\Theta }} = \left\{ {{\bm{q}_u}[n],{\bm{v}_u}[n],{\bm{a}_u}[n],L[n],I[n],\tau [n], d_m[n]} \right\}\ $ and ${{\tilde E}_{UAV}} = d_t\sum\limits_{n = 1}^N {\left[ {{c_1}{{\left\| {{{\bm{v}}_u}[n]} \right\|}^3} + \frac{{{c_2}}}{{\tau [n]}}\left( {1 + \frac{{{{\left\| {{{\bm{a}}_u}[n]} \right\|}^2}}}{{{g^2}}}} \right)} \right]}  + {\Delta _K}$.
Note that  $ - \tilde R_{s,u}^l[n]\ $, ${{\tilde E}_{UAV}}$ and all the constraints are convex. Problem (\ref{eq31}) can thus be solved by employing fractional programming methods, e.g., the Dinkelbach's algorithm, which aims to identify a root of the equation $F(\lambda ) = 0\ $ with updating $\lambda $ in each iteration wherein $F(\lambda )\ $ is given by
\begin{spacing}{0.6}
\begin{equation}\label{eq32}
\mathop {\max }\limits_{\bf{\Theta }} \sum\limits_{n = 1}^N {\tilde R_{s,u}^l[n]}  - \lambda {{\tilde E}_{UAV}}
\end{equation}
\begin{equation*}\label{eq00}
\;\;\;\;\;\;\rm{s.t.}\;\;\;\;(\ref{eq1})-(\ref{eq6}),(\ref{eq17b}),(\ref{eq233}),(\ref{eq30})-(\ref{eq311}),(\ref{eq331}),(\ref{eq3210}).
\end{equation*}
\end{spacing}
\noindent Problem (\ref{eq32}) is a standard convex optimization problem and thus can be solved by the interior-point method \cite{convex}.

As a result, the EE maximum problem is solved as a standard convex optimization problem in an iterative manner with double loops. In the outer loop, by introducing slack variables and SCA technique, we optimize the lower bound of (\ref{eq16}) as (\ref{eq31}) until the fractional increase of the objective function of (\ref{eq31}) is below a small threshold $\mu$. In the inner loop, we solve (\ref{eq32}) with the Dinkelbach's algorithm until the gap between 0 and the value of the objective function is below a small threshold $\eta $. The details of the proposed algorithm are presented in \textbf{Algorithm 1}. It is worth pointing out that \textbf{Algorithm 1} has theoretically provable convergence:

To begin with, one can obtain the gradients of $\tilde R_{s,u}^{}[n]\ $ and $\tilde R_{s,u}^l[n]\ $ with respect to $L_{s,u}^{}[n]\ $ and $I_u^{}[n]\ $ as
\begin{equation}\label{eq35}
\begin{aligned}
{\nabla _{{L_{s,u}}[n],{I_u}[n]}}\tilde R_{s,u}^{}[n]\! = \!& - B{\log _2}e/(L_{s,u}^{}[n] \!+ \!{(L_{s,u}^{}[n])^2}I_u^{}[n])\\
&- B{\log _2}e/(I_u^{}[n] + {(I_u^{}[n])^2}L_{s,u}^{}[n]), \nonumber
\end{aligned}
\end{equation}
and
\begin{equation}\label{eq36}
\begin{aligned}
{\nabla _{{L_{s,u}}[n],{I_u}[n]}}\tilde R_{s,u}^l[n] \!=\!&  - B{\log _2}e/(L_{s,u}^f[n] \!+\! {(L_{s,u}^f[n])^2}I_u^f[n])\\
& - B{\log _2}e/(I_u^f[n] + {(I_u^f[n])^2}L_{s,u}^f[n]). \nonumber
\end{aligned}
\end{equation}

\noindent The gradients of $V_u^{}[n]$ and $V_u^l[n]$ with respect to $ {{\bm{v}}_u}[n]\ $ can be derived as ${\nabla _{{{\bm{v}}_u}[n]}}V_u^{}[n] = 2{\bf{v}}_u^{}[n]$ and ${\nabla _{{{\bm{v}}_u}[n]}}V_u^l[n] = 2{\bf{v}}_u^f[n]$, respectively. Hence, when ${\bm{q}}_u^{}[n]\! =\! {\bm{q}}_u^f[n]$, ${\bm{v}}_u^{}[n]\! = \!{\bm{v}}_u^f[n]$, $L_{s,u}^{}[n] \!= \!L_{s,u}^f[n]$ and $I_u^{}[n] \!= \!I_u^f[n]$, we have $ {\nabla _{{L_{s,u}}[j],{I_u}[n]}}\tilde R_{s,u}^{}[n] = {\nabla _{{L_{s,u}}[j],{I_u}[n]}}\tilde R_{s,u}^l[n]$ and ${\nabla _{{{\bm{v}}_u}[n]}}V_u^{}[n] = {\nabla _{{{\bm{v}}_u}[n]}}V_u^l[n]$. Meanwhile, when ${\bm{q}}_u^{}[n] \!\!\!= \!\!\!{\bm{q}}_u^f[n]$, ${\bm{v}}_u^{}[n]\!\!\! = \!\!\!{\bm{v}}_u^f[n]$, $L_{s,u}^{}[n] \!\!\!= \!\!\! L_{s,u}^f[n]$, and $I_u^{}[n] = I_u^f[n]$, it can be easily observed that all the inequalities in (\ref{eq233}), (\ref{eq30}), (\ref{eq29}), (\ref{eq311}) and (\ref{eq3210}) hold with equality. Thus, \textbf{Algorithm 1} converges to a point satisfying the KKT conditions of the original problem [11, Proposition 3].

Note that each iteration of \textbf{Algorithm 1} requires solving the convex optimization problem (\ref{eq32}) by applying the interior point method, \textbf{Algorithm 1} has a polynomial complexity
${I_i}{I_o}O\left( {{{\left( {10N} \right)}^3}} \right)$ in the worst case, where $10N$ is the number of variables, while ${I_i}$ and
${I_o}$ are the number of inner loop and that of outer loop iterations, respectively.

\vspace{-5pt}
\begin{algorithm}[h]
\caption{Proposed algorithm for solving problem (\ref{eq16})}
\begin{algorithmic}[1]
\STATE{\textbf{Initialization:} Denote ${\bf{\Theta }} $ in ${k_{th}} $ iteration as ${{\bf{\Theta }}^k} $. Initialize a feasible solution as ${{\bf{\Theta }}^0} $. }

\STATE{\textbf{Repeat} (outer loop)}

\STATE{\textbf{Initialization:} Initialize $\lambda $ in ${k_{th}} $ iteration as \quad ${\lambda ^k} = {{\sum\limits_{n = 1}^N {\tilde R_{s,u}^l[n]} } \mathord{\left/
 {\vphantom {{\sum\limits_{n = 1}^N {\tilde R_{s,u}^l[n]} } {{{\tilde E}_{UAV}}}}} \right.
 \kern-\nulldelimiterspace} {{{\tilde E}_{UAV}}}}\ $ and $k=k+1$.}

\STATE{\textbf{Repeat} (inner loop)}

\STATE{Compute ${{\bf{\Theta }}^k}$ via (\ref{eq32}).}

\STATE{Update ${\lambda ^k} = {{\sum\limits_{n = 1}^N {\tilde R_{s,u}^l[n]} } \mathord{\left/
 {\vphantom {{\sum\limits_{n = 1}^N {\tilde R_{s,u}^l[n]} } {{{\tilde E}_{UAV}}}}} \right.
 \kern-\nulldelimiterspace} {{{\tilde E}_{UAV}}}}$}.

\STATE{\textbf{Until} The convergence condition is satisfied.}

\STATE{Update ${\tilde R_{s,u}^l[n]}$ and ${\tilde E_{UAV}}$ in (\ref{eq32}).}

\STATE{\textbf{Until} The convergence condition is satisfied.}
\end{algorithmic}
\end{algorithm}
\vspace{-10pt}
\section{Numerical Results}
\renewcommand{\arraystretch}{0.9}

In this section, numerical results are provided to show the effectiveness of the proposed algorithm (denoted as ``Max EE"). Besides, we adopt two benchmark schemes. Specifically, in the first one, the throughput maximization problem (denoted as ``Max Throughput") is solved by removing the denominator of (\ref{eq31}) as in \cite{Wu2019b}, while in the second one, the EE maximization problem is solved with no jamming signals (denoted as ``Max EE without jamming") as in \cite{Zeng2017a}.

The parameters are set as follows \cite{Zeng2017a}. The time slot length $d_t=0.5$ s. The communication bandwidth is $B=0.1$ MHz. The noise power spectrum density is ${N_0=-169}$ dBm/Hz. Thus, the corresponding noise power is $\sigma ^2=N_0B=-119$ dBm. The transmit power of the source GN $s$ and the jammer $m$ are ${P_s} = 0.1$ W and ${P_m} = 0.1$ W, respectively. Moreover, we set ${c_1}{\rm{  =  9}}{\rm{.26}} \times {\rm{1}}{{\rm{0}}^{ - 4}}$ and ${c_2}{\rm{  =  2250}}$. The altitude of level flight $ H=100$ m. The maximum and minimum speed of UAV are $V_{\max}  = 100$ m/s and $V_{\min} = 3$ m/s, respectively. The maximum acceleration of UAV is ${a_{\max }} = 5 $ m/s$^2$. The channel power gain at the reference distance $d_0=1$ m is $\beta_0=-60$ dB. The convergence threshold for outer loop and inner loop are $\mu  = {10^{ - 3}}$ and $\eta  =  \pm 10$, respectively. The locations of the source GN $s$, the startpoint and the endpoint are set as $(0, 1000, 0)$m, $(-500, 0, H)$m and $(500, 0, H)$m, respectively.

\begin{figure}[!t]
\centering
\subfigure[Case 1] {\includegraphics[width=.48\linewidth]{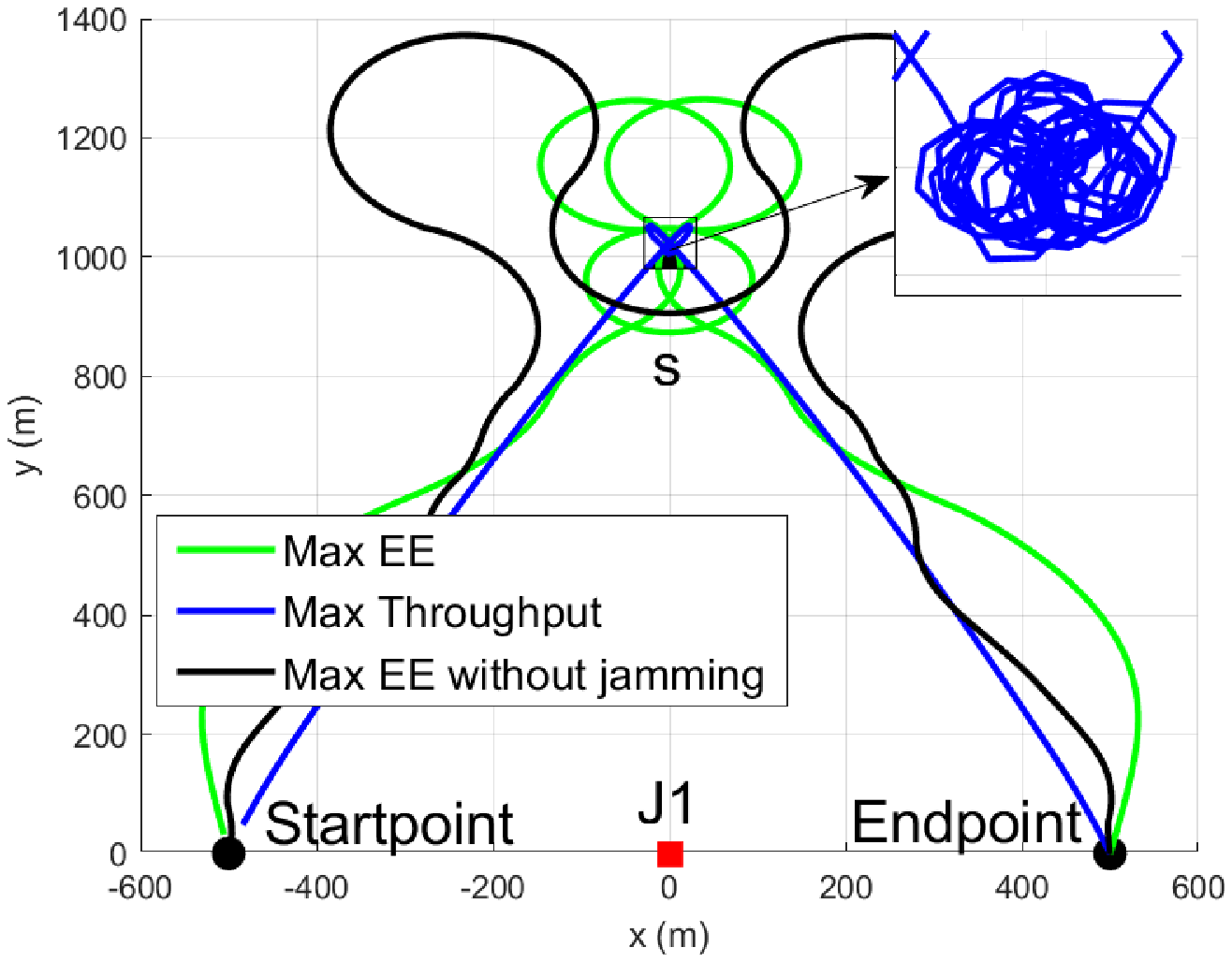}}
\subfigure[Case 2] {\includegraphics[width=.48\linewidth]{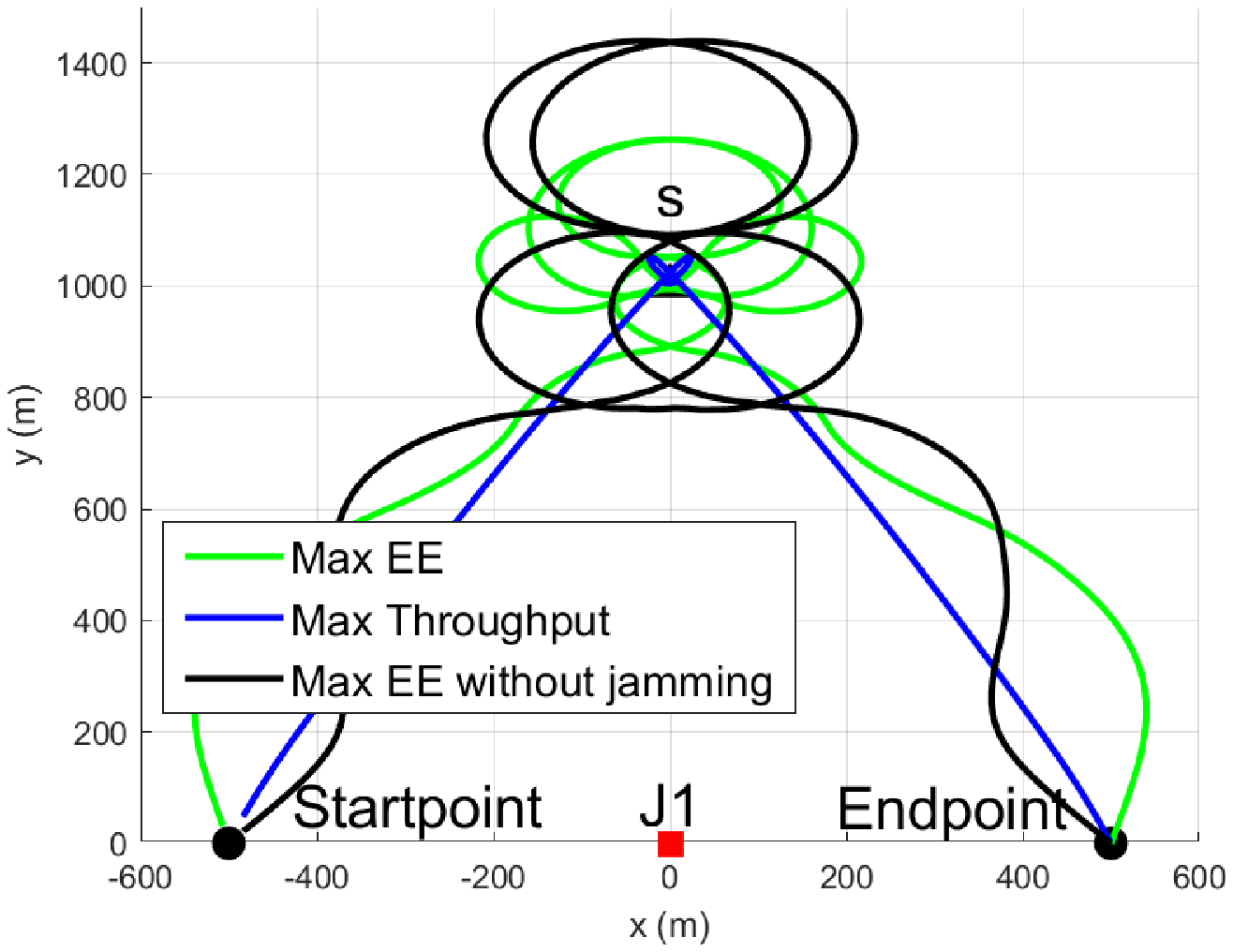}}
\caption{UAV's trajectory of the ``Max EE", ``Max Throughput" and ``Max EE without jamming" algorithms.}\label{12}
\vspace*{-10.0pt}
\end{figure}

In Fig. \ref{12}(a) and Fig. \ref{12}(b), we set one jammer in $(0, 0, 0)$m, and the UAV's trajectories of the ``Max EE", ``Max Throughput" and ``Max EE without jamming" algorithms when $T  =  150$ s and $T  =  200$ s are illustrated as case 1 and case 2, respectively. It is observed that for the ``Max Throughput" algorithm, the UAV tends to hover above $s$ for the maximum possible duration to maintain the best communication channel. Nevertheless, the minimum speed constraint forces the UAV to hover around $s$ instead of hovering still above it. Meanwhile, with the ``Max EE" and ``Max EE without jamming" algorithms, upon approaching $s$, the UAV hovers around following an approximately ``S" shape trajectory. And the lager $T$ is, the more ``S" shape is present, which indicates that such ``S" shape trajectory is expected to maintain a sufficiently good communication channel yet without excessive energy consumption. Moreover, the trajectory of the ``Max EE" algorithm is closer to $s$ than that of the ``Max EE without jamming" algorithm in general for a closer distance to $s$ can reduce the impact of jamming signals.

\begin{figure}[!t]
\centering
\subfigure[Speed of case 1] {\includegraphics[width=.48\linewidth]{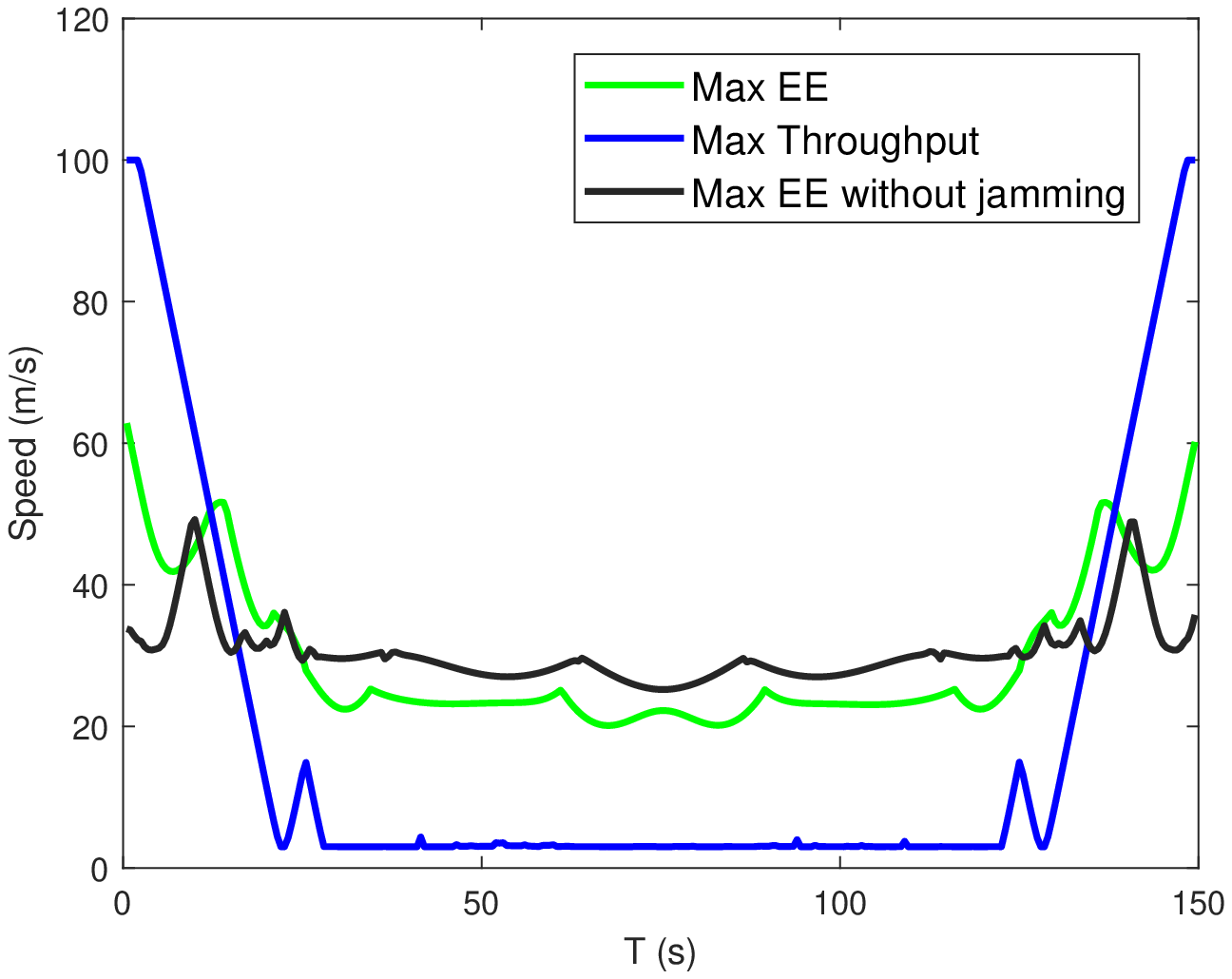}}
\subfigure[Speed of case 2] {\includegraphics[width=.48\linewidth]{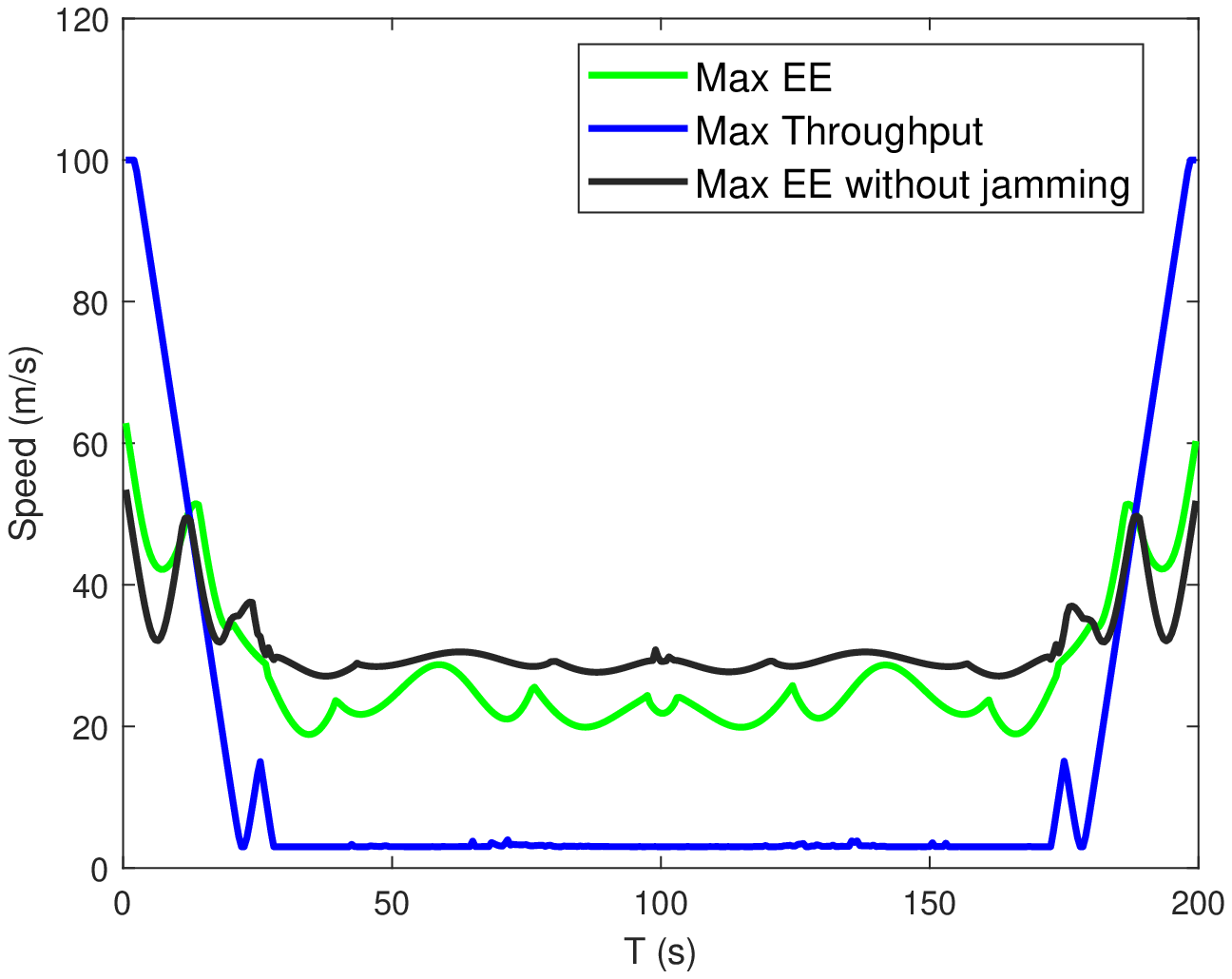}}
\caption{UAV's speed of ``Max EE", ``Max Throughput" and ``Max EE without jamming" algorithms.}\label{12s}
\vspace*{-10.0pt}
\end{figure}

Fig. \ref{12s}(a) and Fig. \ref{12s}(b) further illustrate the speed of the three algorithms in case 1 and case 2, respectively. It is found that for the ``Max Throughput" algorithm, the UAV first flies towards $s$ with high speed, then gradually speeds down to a low speed and hovers around $s$, and finally speeds up to the endpoint. Particularly, in all cases, the time spent on the flight from the startpoint to $s$ and from $s$ to the endpoint is the same. This is because the additional time is used for hovering so as to achieve high throughput. On the other hand, the speed of the ``Max EE" and ``Max EE without jamming" algorithms both fluctuate between the value of 20 and 40 in general, which indicates that flying in such speed is energy efficient.

\begin{figure}[!t]
\centering
\subfigure[Case 3] {\includegraphics[width=.48\linewidth]{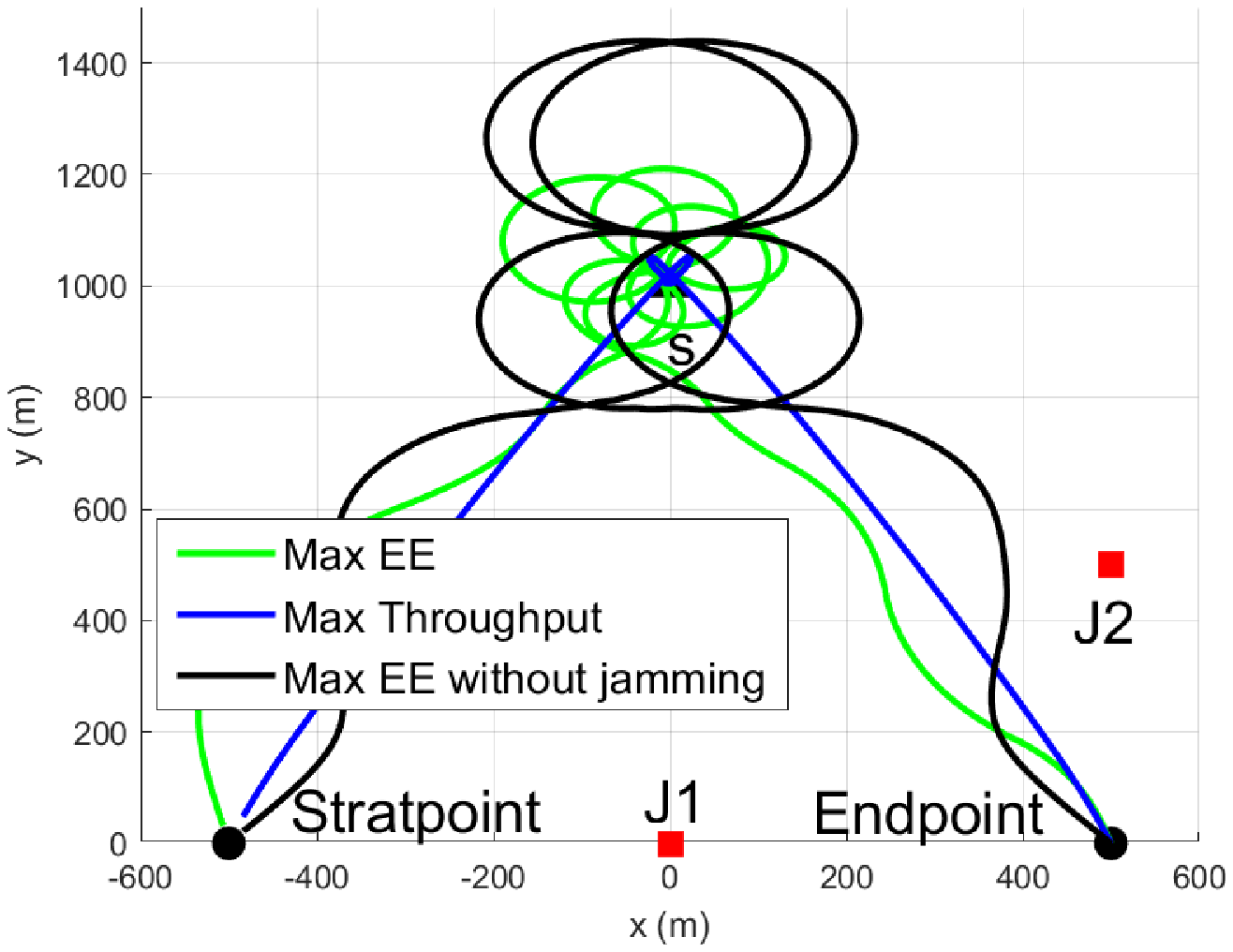}}
\subfigure[Case 4] {\includegraphics[width=.48\linewidth]{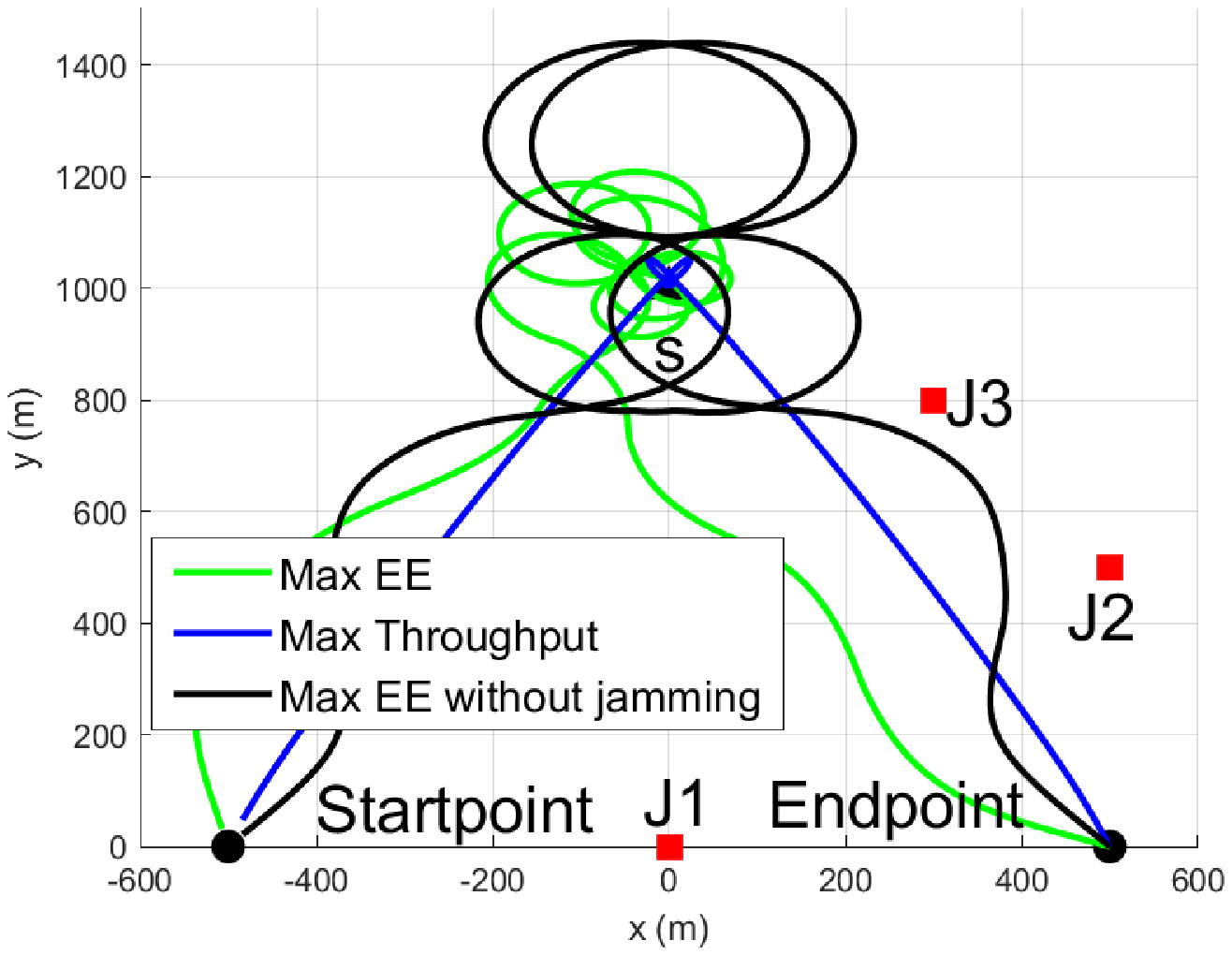}}
\caption{UAV's trajectory of the three algorithms with different jammers' deployment.}\label{34}
\vspace*{-10.0pt}
\end{figure}

Moreover, to verify the proposed ``Max EE" algorithm with multiple jammers, we set $T=200$ s, two jammers in $(0, 0, 0)$m and $(500, 500, 0)$m as case 3 and three jammers in $(0, 0, 0)$m, $(500, 500, 0)$m and $(300, 800, 0)$m as case 4, respectively. As shown in Fig. \ref{34}, the UAV's trajectories of  the ``Max Throughput" and ``Max EE without jamming" algorithms are almost the same. This is because the optimal trajectories for both algorithms are the same, i.e., right above/around $s$ no matter how the jammers are deployed. Meanwhile, for the ``Max EE" algorithm, the successively added jammer in the right hand side of the trajectory forces the trajectories to shift left in general. Hence, it is verified that the proposed ``Max EE" algorithm can adjust UAV's trajectory accordingly in front of various jammers' deployment so as to strike a better balance between throughput and energy consumption.

\begin{table*}[tb]
\centering
\caption{\textcolor[rgb]{0.00,0.00,0.00}{Performance Comparison between Different Algorithms}}
\begin{tabular}{|c|c|c|c|c|c|}
\hline
Algorithm                       & Case  & \begin{tabular}[c]{@{}c@{}}Average\\ Speed\\ $(m/s)$\end{tabular} & \begin{tabular}[c]{@{}c@{}}Sum\\ Throughput\\ $(kbits)$\end{tabular} & \begin{tabular}[c]{@{}c@{}}Energy\\ Consumption\\ $(joule)$\end{tabular} & \begin{tabular}[c]{@{}c@{}}Energy\\ Efficiency\\ $(kbits/joule)$\end{tabular} \\ \hline
\multirow{5}{*}{Max Throughput}
    & 1                                                                         & 18.3                                                               & 172272                                                            & 108332                                                               & 1.59                                                                        \\ \cline{2-6}

& 2                                                                       & 14.5                                                               & 239116                                                            & 152688                                                               & 1.57                                                                        \\ \cline{2-6}
                                & 3                                                                        & 14.4                                                               & 182109                                                             & 155962                                                               & 1.17                                                                        \\ \cline{2-6}
                                & 4                                                                        & 14.4                                                               & 123260                                                             & 156025                                                               & 0.79                                                                         \\  \hline
\multirow{5}{*}{Max EE without jamming}
    & 1                                                                         & 30.5                                                              & 93104                                                            & 18279                                                               & 5.09                                                                        \\ \cline{2-6}

& 2                                                                       & 31.3                                                               & 145874                                                            & 24610                                                               & 5.93                                                                       \\ \cline{2-6}
                                & 3                                                                        & 31.3                                                               & 94050                                                             & 24610                                                               & 3.82                                                                        \\ \cline{2-6}
                                & 4                                                                        & 31.3                                                              & 55770                                                             & 24610                                                               & 2.27                                                                       \\  \hline
\multirow{5}{*}{Max EE}

  & 1                                                                        & 29.7                                                              & 124520                                                             & 20832                                                                  & 5.98                                                                         \\ \cline{2-6}

   & 2                                                                      & 28.1                                                               & 175528                                                             & 27671                                                                  & 6.34                                                                         \\ \cline{2-6}
                                & 3                                                                        & 26                                                               & 133882                                                             & 29149                                                                  & 4.59                                                                        \\ \cline{2-6}
                                & 4                                                                         & 25.7                                                               & 87494                                                              & 29841                                                                  & 2.93                                                                        \\  \hline

\end{tabular}
\end{table*}

To compare the three algorithms more deeply, we show the average speed, sum throughput, energy consumption and EE of cases 1-4 in Table I. It is observed that the proposed ``Max EE" algorithm achieves higher EE than the benchmark algorithms. This is as expected because the ``Max Throughput" algorithm focuses only on maximizing throughput and thus leads to excessively high energy consumption, while the ``Max EE without jamming" algorithm can achieve neither the highest EE nor the highest throughput. Meanwhile, the ``Max Throughput" algorithm has a much lower average speed than the ``Max EE" and ``Max EE without jamming" algorithms in general. This is because the UAV of the ``Max Throughput" algorithm will fly to $s$ as soon as possible with high speed and then slow down to spend most of the time to hover around $s$ with low speed. Moreover, noting that the energy consumptions of the ``Max EE without jamming" algorithm in cases 2-4 are the same, thus the EE of the ``Max EE without jamming" algorithm is only affected by the jammers' deployment.

\begin{figure}[!t]
\centering
\subfigure[EE versus $M$ with $T=200$ s.] {\includegraphics[width=.48\linewidth]{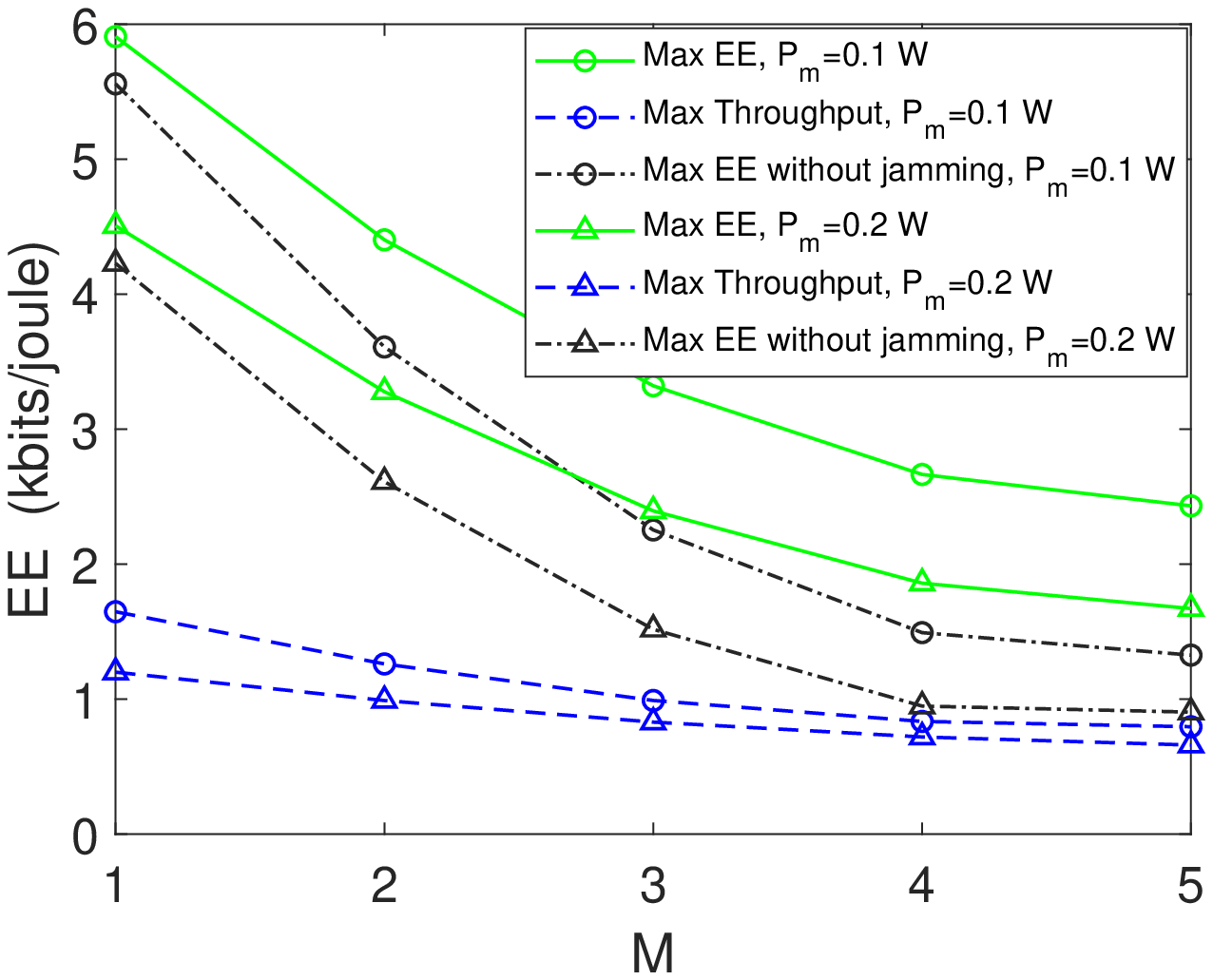}}
\subfigure[EE versus $T$ with $M=1$.] {\includegraphics[width=.48\linewidth]{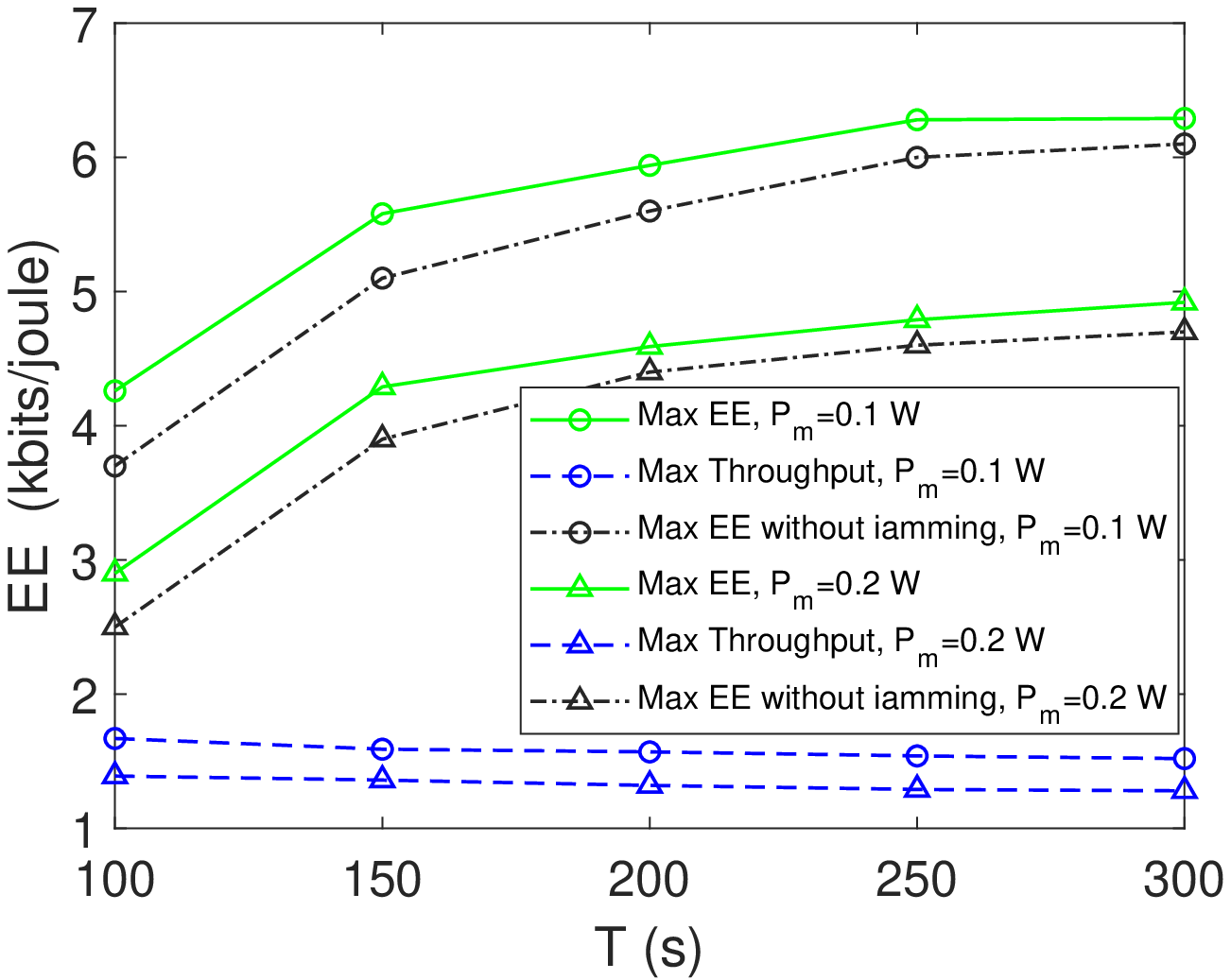}}
\caption{EE of UAV in different algorithms.}\label{2VS}
\vspace*{-10.0pt}
\end{figure}

Without loss of generality, we set a square whose corners are set as $(-500, 0, 0)$m, $(500, 0, 0)$m, $(-500, 1000, 0)$m and $(500, 1000, 0)$m, and generate the jammers randomly inside it. The EE of the three algorithms versus jammers' number $M$ and flight time $T$ are illustrated in Fig. \ref{2VS}(a) and Fig. \ref{2VS}(b), respectively. In Fig. \ref{2VS}(a), it can be observed that the ``Max EE" algorithm always performs better than the ``Max Throughput" and ``Max EE without jamming" algorithms. However, with the increase of $M$, the EE of the ``Max Throughput" algorithm decreases relatively slower than the ``Max EE" algorithm. This is as expected because as observed in Table I, the energy consumption of both algorithms keep relatively stable as $M$ increases and that of the ``Max EE" is much lower than that of the ``Max Throughput", thus the similar reduction of throughput has a larger impact on the EE of the "Max EE" algorithm. On the contrary, the EE of the ``Max EE without jamming" algorithm decreases relatively faster than the ``Max EE" algorithm. This is because the ``Max EE without jamming" algorithm can not reduce the impact of jamming signals and thus strong jamming signals will lead to poor system performance. In Fig. \ref{2VS}(b), it is observed that the gap between EE of the ``Max EE" algorithm and that of the benchmark algorithms become stable with growing $T$, which indicates that with sufficient flight time, the ``Max EE" algorithm can lead to relative ideal EE.


\vspace{-5pt}

\section{Conclusion}

In this letter, the EE of the UAV-enabled communication in the presence of malicious jammers was studied by considering the propulsion energy consumption. With the aid of SCA technique and Dinkelbach's algorithm, an iterative algorithm was proposed to solve the formulated non-convex problem. Numerical results showed that the proposed algorithm outperformed the benchmark algorithms significantly especially when the flight time was sufficiently long and can adjust the trajectory in front of various jammer's deployments. It is very interesting to consider a more general channel model that contains both LoS and NLoS links. However, such a model will lead to a totally new and sophisticated optimization problem and the solution for it is nontrivial, which will be left as our future work.


\vspace{-5pt}
\begin{thebibliography}{99}

\bibitem{Zeng2016}
Y. Zeng, Q. Wu, and R. Zhang,``{Accessing from the sky: A tutorial on UAV communications for 5G and beyond},'' \emph{Proc. IEEE},
 vol. 107, no. 12, pp. 2327-2375, Dec. 2019.


\bibitem{WQ}
  Q. Wu, Y. Zeng, and R. Zhang, ``{Joint trajectory and communication design for multi-UAV enabled wireless networks},'' \emph{IEEE Trans. Wirel. Commun.}, vol. 17, no. 3, pp. 2109-2121, Mar. 2018.

\bibitem{xh}
S. Li, B. Duo, X. Yuan, Y. Liang and M. Di Renzo, ``{Reconfigurable intelligent surface assisted UAV communication: Joint trajectory design and passive beamforming},'' \emph{IEEE Wirel. Commun. Lett.}, vol. 9, no. 5, pp. 716-720, May. 2020.



\bibitem{Zeng2017a}
Y. Zeng and R. Zhang, ``{Energy-efficient UAV communication with trajectory optimization},'' \emph{IEEE Trans. Wirel. Commun.}, vol.~16, no.~6, pp.3747-3760, Jun. 2017.


\bibitem{Xiao}
L.~Xiao, Y.~Xu, D.~Yang, and Z.~Yong, ``{Secrecy energy efficiency maximization for
  UAV-enabled mobile relaying},'' \emph{IEEE Trans. Green Commun. Netw.}, vol. 4, no. 1, pp. 180-193, Mar. 2020.

\bibitem{wqq}
Q. Wu, W. Mei, and R. Zhang, ``{Safeguarding wireless network with
UAVs: A physical layer security perspective},'' \emph{IEEE Wirel. Commun.}, vol. 26, no. 5, pp. 12-18, Oct. 2019.



\bibitem{Wu2019b}
Y.~Wu, W. Fan, W.~Yang, X.~Sun, and X. Guan, ``{Robust trajectory and communication design for multi-UAV enabled wireless networks in the presence of jammers},'' \emph{IEEE Access}, vol. 8, pp. 2893-2905, 2020.



 \bibitem{Wang2018g}
H.~Wang, J.~Wang, G.~Ding, J.~Chen, Y.~Li, and Z.~Han, ``{Spectrum sharing
  planning for full-duplex UAV relaying systems with underlaid D2D
  communications},'' \emph{IEEE J. Sel. Areas Commun.}, vol.~36, no.~9, pp.
  1986--1999, Sep. 2018.




\bibitem{x}
X. Linet al., ``{The sky is not the limit: LTE for unmanned aerial vehicles},'' \emph{IEEE Commun. Mag.}, vol. 56, no. 4, pp. 204-210, Apr. 2018.





\bibitem{convex}
S.~Boyd and L.~Vandenberghe, \emph{Convex Optimization}. Cambridge, U.K.:
Cambridge Univ. Press, 2004.


\bibitem{kkt}
A. Zappone, E. Bjornson, L. Sanguinetti, and
E. Jorswieck, ``{Globally optimal energy-efficient power control and receiver design in wireless networks},'' \emph{IEEE Trans. Signal Process.}, vol. 65, no.11, pp. 2844-2859,  Jun. 2017.
\end{thebibliography}
\end{document}